\documentclass{twocolumns}
\usepackage{natbib}
\newcommand{\etal}   {{\rm ~et al.}}

\shortauthors{Kondratko\etal} \shorttitle{Discovery of Water Maser Emission in
Eight AGN with the DSN}

\begin{document}

\title{Discovery of Water Maser Emission in Eight AGN with $70$-m Antennas of
NASA's Deep Space Network}

\author{
P. T. Kondratko,\altaffilmark{1} L. J. Greenhill,\altaffilmark{1} J. M.
Moran,\altaffilmark{1} J. E. J. Lovell,\altaffilmark{2} T. B. H.
Kuiper,\altaffilmark{3} D. L. Jauncey,\altaffilmark{2}}
\author{ L. B. Cameron,\altaffilmark{4}
J. F. G\'omez,\altaffilmark{5} C. Garc\'{\i}a-Mir\'o,\altaffilmark{6} E.
Moll,\altaffilmark{6} I. de Gregorio-Monsalvo,\altaffilmark{7} and E.
Jim\'enez-Bail\'on\altaffilmark{7}}

\altaffiltext{1}{Harvard-Smithsonian Center for Astrophysics, 60 Garden St,
Cambridge, MA 02138 USA}

\altaffiltext{2}{Australia Telescope National Facility, CSIRO, Epping, NSW 2121
Australia}

\altaffiltext{3}{Jet Propulsion Laboratory, 4800 Oak Grove Dr, Pasadena, CA 91109
USA}

\altaffiltext{4}{Raytheon, Canberra Deep Space Communications Complex,
Tidbinbilla Rd, Paddy's Creek, ACT 2609, Australia}

\altaffiltext{5}{Instituto de Astrof\'{\i}sica de Andaluc\'{\i}a, CSIC, Apartado
3004, E-18080 Granada, Spain}

\altaffiltext{6}{Madrid Deep Space Communication Complex, INTA-NASA, Paseo del
Pintor Rosales 34-bajo, E-28008 Madrid, Spain}

\altaffiltext{7}{Laboratorio de Astrof\'{\i}sica Espacial y F\'{\i}sica
Fundamental, INTA, Apartado 50727, E-28080 Madrid, Spain }

\email{pkondrat@cfa.harvard.edu}

\begin{abstract}
We report the discovery of water maser emission in eight active galactic nuclei
(AGN) with the \hbox{$70$-m} NASA Deep Space Network (DSN) antennas at
Tidbinbilla, Australia and Robledo, Spain. The positions of the newly discovered
masers, measured with the VLA, are consistent with the optical positions of the
host nuclei to within $1\sigma$ ($0\rlap{.}''3$ radio and $1\rlap{.}''3$ optical)
and most likely mark the locations of the embedded central engines. The spectra
of two sources, NGC\,3393 and NGC\,5495, display the characteristic spectral
signature of emission from an edge-on accretion disk, with orbital velocities of
$\sim 600$ and $\sim 400$\,km\,s$^{-1}$, respectively. In a survey with DSN
facilities of $630$ AGN selected from the NASA Extragalactic Database, we have
discovered a total of $15$ water maser sources. The resulting incidence rate of
maser emission among nearby ($v_{sys} < 7000$ km s$^{-1}$) Seyfert\,$1.8-2.0$ and
LINER systems is $\sim10\%$ for a typical rms noise level of $\sim14$\,mJy over
$1.3$\,km\,s$^{-1}$ spectral channels. As a result of this work, the number of
nearby AGN ($v_{sys} < 7000$ km s$^{-1}$) observed with $<20$\,mJy rms noise has
increased from $130$ to $449$.
\end{abstract}


\section{Introduction}

Water maser emission ($\lambda=1.3$ cm) is currently the only resolvable tracer
of warm dense molecular gas in the inner parsec of active galactic nuclei (AGN)
and has been detected to-date in approximately $60$ nuclei. In eight water maser
systems that have been studied with sub-milliarcsecond resolution using Very Long
Baseline Interferometry (VLBI), the mapped emission appears to trace structure
and dynamics of molecular disks $0.1$ to $1$ pc from supermassive black holes:
Circinus (Greenhill et al. 2003), NGC\,1068 \citep{Greenhill1997c}, NGC\,4258
\citep{Miyoshi1995}, NGC\,3079 \citep{Trotter1998}, IC\,2560
\citep{Ishihara2001}, NGC\,5793 \citep{Hagiwara2001b}, NGC\,4945
\citep{Greenhill1997}, NGC\,1386 \citep{Braatz1997AAS}. Due to spatial coherence
in line-of-sight velocity within the rotating structures, maser emission is
detected preferentially in edge-on disks along the midline (i.e., the diameter
perpendicular to the line of sight) and close to the line of sight towards the
center. A characteristic spectral signature of emission from an edge-on disk thus
consists of a spectral-line complex in the vicinity of the systemic velocity
(low-velocity emission) and two spectral-line complexes symmetrically offset from
the systemic velocity by the orbital velocity of the disk (high-velocity
emission). Sources that display such spectra are referred to here as
high-velocity systems and constitute $\sim40\%$ of the known nuclear water
masers.

The study of high-velocity systems with VLBI has significantly contributed to our
understanding of the immediate vicinity (i.e., $\lesssim1$\,pc) of supermassive
black holes. In three of these systems
--- NGC\,4258 \citep{Miyoshi1995}, NGC\,1068 \citep{Greenhill1997c}, and the Circinus
Galaxy (Greenhill et al. 2003)\nocite{Greenhill2003} --- resolved position and
line-of-sight velocity data provided evidence for differential rotation and
enabled accurate estimation of black hole mass and pc-scale molecular disk
structure. Another system, NGC\,3079, in which the rotation curve traced by the
maser emission appears flat, was interpreted in the context of a pc-scale, thick,
edge-on, self-gravitating, and possibly star forming molecular disk
\citep{Kondratko2005, Yamauchi2004}. In addition to mapping pc-scale molecular
disk structure and accurately weighing supermassive black holes, nuclear water
maser emission has also been used as a distance indicator. Distance determination
is possible for systems where a robust pc-scale disk model from VLBI maps is
combined with a measurement of either maser proper motions or drifts in
line-of-sight velocity of spectral features (i.e., centripetal acceleration). The
distance to NGC\,4258 obtained in this manner is the most accurate extragalactic
distance thus far, is independent of standard candle calibrators such as Cepheids
(Herrnstein et al. 1999)\nocite{Herrnstein1999}, and it has contributed to
analysis of the Cepheid period-luminosity relation \citep[Freedman et al.
2001;][]{Newman2001}\nocite{Freedman2001}.

The $\sim60$ known nuclear water masers are in the great majority of cases
associated with Seyfert\,$1.8-2.0$ or LINER nuclei; only five systems are
exceptions: NGC\,5506 \citep[NLSy1 from][]{Nagar2002,Braatz1996}, NGC\,4051
\citep[Sy1.5 from the NED;][]{Hagiwara2003}, NGC\,2782 \citep[Sy1, starburst from
the NED;][]{Braatz2004}, NGC\,4151 \citep[Sy1.5 from the NED;][]{Braatz2004}, and
3C\,403 \citep[FRII;][]{Tarchi2003}. In the context of the unified AGN model,
Seyfert\,$1.8-2.0$ systems contain an active nucleus and an obscuring structure
along the line of sight to the central engine \citep{Lawrence1982,
Antonucci1993}. Irradiation of molecular gas by X-rays from the central engine is
a plausible means of exciting maser emission \citep*[e.g.,][]{Neufeld1994}, which
might explain the association of maser emission with nuclear activity in general.
Maser emission might be associated with Seyfert\,$1.8-2.0$ systems in particular
because, over a range of AGN luminosity, the shielding column density that
provides the obscuring geometry maintains not only a reservoir of molecular gas
but also physical conditions conducive to maser action, which are temperatures of
$250-1000$\,K and H$_2$ number densities of $10^{8-10}$\,cm$^{-3}$
\citep*[e.g.,][]{Desch1998}. Since there is good evidence for a physical
relationship between LINER systems and the AGN phenomenon
\citep[e.g.,][]{Ho1997,Ho1999,Ho2003}, the distinction between these two types of
activity in the context of maser emission might simply be one of luminosity,
whereby a large fraction of LINERs correspond to low-luminosity analogues of
Seyfert\,$1.8-2.0$ systems.

Since water maser emission is typically weak ($\ll 0.1$\,Jy) and the velocity
range of emission is determined by the orbital velocity of the molecular disk
($\gg 100$\,km\,s$^{-1}$), surveys designed to detect new water maser sources
require both large, sensitive apertures and wide bandwidth spectrometers. The
detection rate of water maser emission among Seyfert\,$2$ and LINER galaxies with
$v_{sys} < 7000$\,km\,s$^{-1}$ is $\sim4\%$ for surveys with a typical rms noise
level on each source of $\sim60$\,mJy \citep[thereafter,
sensitivity;][]{Braatz1996}. This low detection rate, the weakness of the
emission, and its potential wide velocity range make the discovery of new water
maser sources challenging. Most past survey work has been characterized by
limited sensitivity ($\sim60$\,mJy) and narrow bandwidths ($\sim
700$\,km\,s$^{-1}$) and thus might have missed new maser sources because the
emission was either too weak or outside the observing bands. These limitations of
the previous surveys provide the major impetus for the present work. We note that
a recent survey with $\sim3$\,mJy sensitivity (converted to $1.3$\,km\,s$^{-1}$
spectral channels) using the Green Bank Telescope (GBT) has yielded a detection
rate of $\sim20\%$ among Seyfert\,$2$ and LINER systems with $v_{sys} <
7500$\,km\,s$^{-1}$, though it was limited to just $145$ sources
\citep{Braatz2004}.

In order to discover more high-velocity systems, we procured a custom-built
$4096$-channel spectrometer with $5300$\,km\,s$^{-1}$ bandwidth and are
conducting a survey with the \hbox{$70$-m} NASA Deep Space Network (DSN) antennas
at Tidbinbilla, Australia and at Robledo, Spain. We selected our sample from
among $1150$\,AGN with $v_{sys} < 14600$\,km\,s$^{-1}$ listed in the NASA
Extragalactic Database (NED), with preference for Seyfert\,$1.8-2.0$ and LINER
systems at lower recessional velocities. Thus far, we have discovered water maser
emission in $15$ AGN. The first seven discoveries were reported in
\cite{Greenhill2003survey}; here, we present spectra of eight most recent
detections.

\section{Observations}

The discoveries reported here were obtained during the 2004-2005 northern winter
with the Robledo \hbox{$70$-m} antenna and during the 2003 and 2005 southern
winters with the Tidbinbilla \hbox{$70$-m} antenna. The observing system at
Tidbinbilla and its calibration was described in \cite{Greenhill2003survey}. The
observing system setup at Robledo was identical to that at Tidbinbilla. We
estimated the gain curve and aperture efficiency of the Robledo antenna through
measurement of opacity corrected antenna temperature of 3C\,147, for which we
adopted a flux density of $1.82$\,Jy at $22.175$\,GHz \citep{Baars1977}. The
resulting peak efficiency was $0.43\pm 0.09$ at $43\degr\pm 5\degr$ elevation,
which yields a sensitivity of $1.7\pm 0.3$\,$\mbox{Jy}\,\mbox{K}^{-1}$.

To obtain single-polarization total-power spectra of each source, we moved the
telescope every $30$ or $45$\,s between the target source and a reference
position on the sky $\sim 0.2\degr$ away. Antenna rms pointing errors were
typically $7''$ and system temperatures ranged from $40$ to $75$\,K depending on
elevation and weather. Typical $1\,\sigma$ noise levels attained in an
integration time (on+off\,source) of one hour were $\sim 8-17$\,mJy in a
$1.3$\,km\,s$^{-1}$ channel. The spectra reported here have been Hanning smoothed
to an effective resolution of $3.5$\,km\,s$^{-1}$ and corrected for atmospheric
opacity estimated from tipping scans.

\section{Results}

\begin{figure*}[!h]
\epsscale{1.05} \plotone{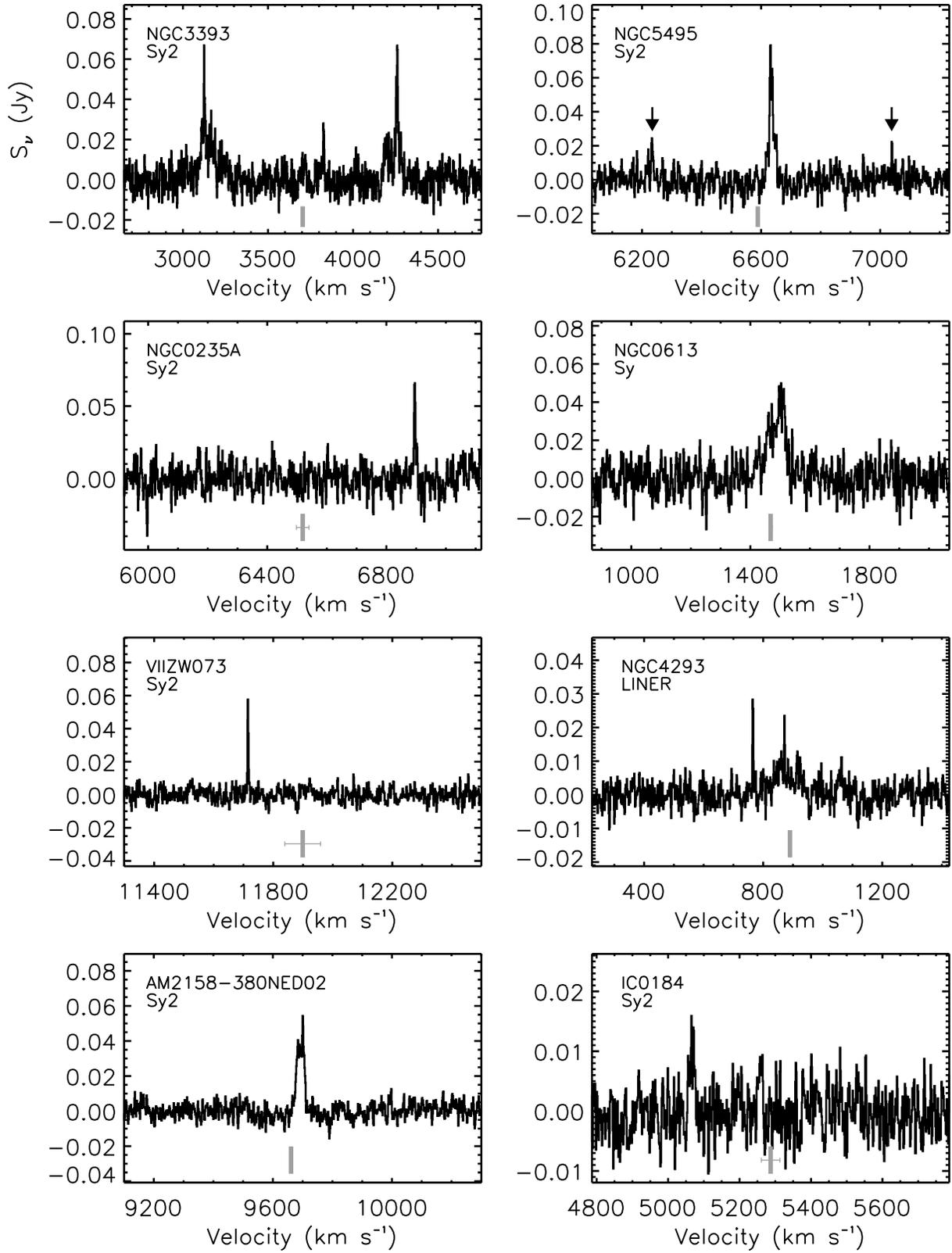} \caption{Spectra of NGC\,3393, NGC\,5495,
NGC\,0235A, NGC\,0613, VII\,ZW\,073, NGC\,4293, AM\,2158-380\,NED02, and IC\,0184
obtained with the \hbox{$70$-m} Deep Space Network antennas at Tidbinbilla,
Australia and Robledo, Spain. The arrows indicate NGC\,5495 high-velocity
features confirmed with the GBT. The velocity was computed in accordance with the
radio definition of Doppler Shift and in the heliocentric reference frame.
Vertical bars depict systemic velocities of the host galaxies and respective
uncertainties.\label{figure}}
        \hrulefill\
\end{figure*}

During the 2003 and 2005 southern winters, we detected maser emission in six AGN
with the Tidbinbilla antenna: AM\,2158-380\,NED02, IC\,0184, NGC\,0235A,
NGC\,0613, NGC\,3393, and NGC\,5495 (Fig. \ref{figure} and Table \ref{Table}).
Maser emission from NGC\,4293 and VII\,ZW\,073 was detected during the 2004-2005
northern winter with the Robledo antenna. Each discovery was confirmed by at
least one observation with another instrument (Table \ref{Table}). Positions of
the maser emission measured with the Very Large Array (VLA) of the
NRAO\footnote{The National Radio Astronomy Observatory is operated by Associated
Universities, Inc., under cooperative agreement with the National Science
Foundation} are consistent with optical positions of the AGN to within $1\sigma$
($0\rlap{.}''3$ radio and $1\rlap{.}''3$ optical; Table \ref{Table}), which
confirms the association of the discovered emission with nuclear activity.

The nuclei that are host to the detected maser emission are spectroscopically
classified as Seyfert\,$2$ or LINER in all but two cases (Table \ref{Table}).
Ambiguity remains in the case of NGC\,0613, which is listed as a possible Seyfert
by \cite{Veron1986}, and in the case of NGC\,0235A, which is classified as
Seyfert\,$2$ by \cite{Monk1986} but as Seyfert\,$1$ by \cite{Maia1987}. There is
some indication of a broad component in H$\alpha$ in the optical spectrum
presented by \cite{Maia1987} (the \cite{Monk1986} spectrum does not cover the
wavelength of H$\alpha$). However, optical spectra presented by \cite{Monk1986}
and \cite{Maia1987} exhibit a narrow H$\beta$ line and a strong
[O\,III]$\lambda$\,$5007$ line relative to H$\beta$. According to the
quantitative classification scheme of \cite{Winkler1992} \citep[see
also][]{Osterbrock1977,Osterbrock1981}, NGC\,0235A thus harbors either a
Seyfert\,$1.9$ or a Seyfert\,$2$ nucleus. A spectrum obtained independently by J.
Huchra (2004, private communication) confirms this classification. It is thus
likely that NGC\,0235A was misclassified based on the hint of broad H$\alpha$
reported by \cite{Maia1987}.

The maser spectrum of NGC\,3393 shows a characteristic spectral signature of
emission from an edge-on disk: two high-velocity complexes ($\sim 70$\,mJy)
symmetrically offset by $\sim 600$\,km\,s$^{-1}$ from the systemic velocity and a
single spectral complex ($\sim 28$\,mJy) within $120$\,km\,s$^{-1}$ of the
systemic velocity. We have confirmed the weak systemic feature with the
Tidbinbilla antenna ($1\sigma = 7$\,mJy) and with the GBT ($1\sigma = 2$\,mJy in
a $0.33$\,km\,s$^{-1}$ spectral channel). If the high-velocity emission indeed
originates from the midline of an edge-on disk, then the orbital velocity of the
disk as traced by the maser emission is $\sim 600$\,km\,s$^{-1}$, making
NGC\,3393 the third fastest known rotator after NGC\,4258 \citep[$\sim
1200$\,km\,s$^{-1}$;][]{Miyoshi1995} and ESO269-G012 \citep[$\sim
650$\,km\,s$^{-1}$;][]{Greenhill2003survey}. Because the high-velocity emission
extends over $\Delta v\sim 200$\,km\,s$^{-1}$, it must occupy a fractional range
of radii $\Delta r/R \approx 2\Delta v/v = 2/3$, which yields $\Delta r\sim
0.15-0.57$\,pc, assuming Keplerian rotation ($v\propto r^{-0.5}$) and a range in
disk sizes from that of NGC\,4258 ($0.16-0.28$\,pc) to NGC\,1068 ($0.6-1.1$\,pc).
The corresponding central mass would be on the order of $5\times 10^7\,M_{\odot}$
and the anticipated centripetal acceleration
--- that is the secular velocity drift of the systemic feature --- would be on
the order of $1$\,km\,s$^{-1}$\,yr$^{-1}$, which should be readily detectable
within one year using single dish monitoring.

We confirmed the detection of high-velocity lines in NGC\,5495 (refer to Fig.
\ref{figure}) with the GBT ($1\sigma = 5$\,mJy in a $0.33$\,km\,s$^{-1}$ spectral
channel). If we again assume that the high-velocity emission originates from the
midline of an edge-on disk, then the orbital velocity of the accretion disk is
$\sim 400$\,km\,s$^{-1}$ while the corresponding central mass and centripetal
acceleration are on the order of $10^7\,M_{\odot}$ and
$0.5$\,km\,s$^{-1}$\,yr$^{-1}$, respectively, assuming disk sizes as in NGC\,4258
and NGC\,1068.

Each of the remaining six detections --- AM\,2158-380\,NED02, IC\,0184,
NGC\,0235A, NGC\,0613, NGC\,4293, and VII\,ZW\,073 --- displays only a single
complex of spectral features. In particular, the spectrum of NGC\,0613 reveals a
very broad emission feature (full width at half-maximum of $\sim 87$ km s$^{-1}$)
and such broad lines have been typically associated with radio jets rather than
molecular disks \citep[e.g.,][]{Peck2003,Claussen1998}. The spectral lines in
NGC\,0235A, VII\,ZW\,073, and IC\,0184 are significantly displaced from systemic
velocities of the host galaxies (by $\sim380$, $\sim180$, and
$\sim220$\,km\,s$^{-1}$, respectively) and thus might be either emission features
associated with radio jets \citep[e.g.,][]{Peck2003,Claussen1998} or
high-velocity emission lines from only a single side of an edge-on accretion
disk. NGC\,0613, NGC\,3393, NGC\,4293, and IC\,0184 have been targeted but not
detected in previous surveys for maser emission. For NGC\,0613, \cite{Braatz1996}
and \cite{Henkel1984} report $1\sigma$ noise levels of $364$\,mJy and $70$\,mJy
in $1.7$ and $1.3$\,km\,s$^{-1}$ spectral channels, respectively, corresponding
to signal-to-noise ratios of $\lesssim 1$ at the current line strength (peak flux
of $\sim 70$\,mJy in a $1.3$ km s$^{-1}$ spectral channel). Considering the large
line-width, broader spectral averaging could have been applied to
\cite{Henkel1984} data to achieve a marginal detection, and it is thus unclear
whether this observation indicates line variability analogous to that observed in
other masers associated with jet activity \citep[e.g.,][]{Peck2003}. For
NGC\,3393, \cite{Braatz1996} report $1\sigma$ noise level of $11$\,mJy in
$0.66$\,km\,s$^{-1}$ channels, which should have been sufficient to detect the
maser emission at its present strength of $\sim 80$\,mJy (in a
$1.3$\,km\,s$^{-1}$ channel). In the case of NGC\,4293, J.\,A.\,Braatz (private
communication) obtained in 1998 a $1\sigma$ noise level in a $1.3$\,km\,s$^{-1}$
channel of $35$\,mJy, which would not have been sufficient to detect the weak
emission (peak of $\sim 40$\,mJy in a $1.3$\,km\,s$^{-1}$ channel). For IC\,0184,
\cite{Braatz1996} report $1\sigma$ noise level of $63$\,mJy in
$0.66$\,km\,s$^{-1}$ channels, again not sufficient to detect the maser at
present line strength (peak of $\sim 27$\,mJy in a $1.3$\,km\,s$^{-1}$ channel).

In a survey with the Tidbinbilla and Robledo antennas of $630$ AGN with $v_{sys}
< 14600$\,km\,s$^{-1}$ selected from the NED (Table \ref{Survey}), we have
detected to-date $15$ new water maser sources \citep[this paper
and][]{Greenhill2003survey}. Since previous searches for water maser emission
reported detections mostly in Seyfert\,$1.8-2.0$ or LINER systems
\cite[e.g.,][]{Braatz1996}, we have focused our survey primarily on obscured
nuclei ($488$\,Seyfert\,$1.8-2.0$ or LINER systems, among which there are the
$15$ new maser sources, and $143$\,Seyfert\,$1.0-1.5$ systems, for which we
report no new detections). Our survey includes $55\%$ of known AGN with
$v_{sys}<14600$\,km\,s$^{-1}$ ($630$ out of $1150$) and is nearly complete to
$7000$\,km\,s$^{-1}$ with $82\%$ of Seyfert\,$1.8-2.0$ and LINER systems ($325$
out of $398$) already observed, although AGN catalogs consolidated by the NED may
themselves be incomplete. The detection rate among nearby ($v_{sys} <
7000$\,km\,s$^{-1}$) Seyfert\,$1.8-2.0$ and LINER galaxies is $4\%$. The
detection rate among nearby Seyfert\,$1.8-2.0$ and LINER galaxies that have been
observed with absolute sensitivity (i.e., $1\sigma$ noise level in flux density
units multiplied by the spectral channel width and converted to luminosity using
$H_o=75$\,km\,s$^{-1}$\,Mpc) of better than $2\,L_{\odot}$ is $4\%$, consistent
with the analogous detection rate of $6\%$ reported by \cite{Braatz1996}.
However, taking into account all known maser sources with peak flux densities
above four times the typical rms achieved in this survey ($4\sigma=56$\,mJy or
$15\,L_{\odot}$ in a $1.3$\,km\,s$^{-1}$ spectral channel at
$v_{sys}=7000$\,km\,s$^{-1}$), the resulting incidence rate of maser emission
among nearby Seyfert\,$1.8-2.0$ and LINER systems is $\sim10\%$, which should be
compared to the analogous incidence rates of $\sim7\%$ and $\sim20\%$ for surveys
with $\sim60$\,mJy \citep[$\sim1$\,km\,s$^{-1}$ spectral channels;][]{Braatz1996}
and $\sim3$\,mJy \citep[converted to $1.3$\,km\,s$^{-1}$ spectral
channels;][]{Braatz2004} sensitivities, respectively. Six of the detections
obtained with DSN (NGC\,0613, NGC\,3393, NGC\,4293, NGC\,5643, NGC\,6300, and
IC\,0184) lie in nuclei that had been targeted in previous surveys, which
demonstrates the importance of either survey sensitivity or source variability.
As a result of this work, the number of nearby AGN ($v_{sys} < 7000$ km s$^{-1}$)
observed with $<20$\,mJy sensitivity has increased from $130$ to $449$.

\section{Future Prospects}

Four of the fifteen detections obtained using the \hbox{$70$-m} antennas of the
DSN
--- NGC\,3393, NGC\,5495 (this study), ESO\,269-G012, and possibly NGC\,6926
\citep{Greenhill2003survey} --- display the archetypal spectral signature of
emission from an edge-on disk. The remaining 11 sources appear to be
non-high-velocity systems, but monitoring and deep integrations (as with the GBT,
which is roughly an order of magnitude more sensitive than the DSN antennas) may
result in detection of high-velocity features. All high-velocity systems stronger
than few mJy are good candidates for follow-up high angular resolution study with
VLBI, where pc-scale molecular disk geometry and black hole mass may be inferred
directly from resolved position and line-of-sight velocity data with high
accuracy and relatively few sources of systematic uncertainty. Assuming disk
sizes of $0.1$ to $1$\,pc (i.e., comparable to NGC\,4258 and NGC\,1068,
respectively) and a VLBI resolution element of $\sim0.3$\,mas, the source
structure should be readily resolved because the AGN are relatively nearby
($3704$\,km\,s$^{-1}<v_{sys}<6589$\,km\,s$^{-1}$). The additional detection with
single dish spectroscopic monitoring of secular drift in the velocities of
low-velocity Doppler components (i.e., centripetal acceleration) may also enable
estimation of geometric distances, as has been done for NGC\,4258 (Herrnstein et
al. 1999). A geometric distance to any of these systems would be of considerable
value since it might help in calibration of the Hubble relation independent of
standard candles such as Cepheids.

We are grateful for the invaluable support provided by the management,
operations, and technical staff of the Canberra and Madrid Deep Space
Communications Complexes. We thank M. Franco for Radio Frequency support and most
especially P. Wolken and the staff charged with scheduling antenna time, without
whose help and guidance this project would have been impossible. This research
has made an extensive use of the NASA/IPAC Extragalactic Database (NED) which is
operated by the Jet Propulsion Laboratory (JPL), California Institute of
Technology, under contract with NASA. This work was supported in part by R\&D
funds of the Smithsonian Astrophysical Observatory. IdG and JFG are partially
supported by grant AYA2002-00376 (FEDER funds) of the Ministerio de Educaci\'on y
Ciencia (Spain). Canberra Deep Space Communication Complex is managed by
Commonwealth Scientific and Industrial Research Organisation (CSIRO) on behalf of
JPL for NASA, with operations and maintenance undertaken by Raytheon, Australia.
The Australia Telescope National Facility is funded by the Commonwealth
Government for operation as a national facility by the CSIRO.

\bibliography{ms}

\clearpage

\clearpage

\begin{deluxetable}{llrrrcrrrl}
%\tablewidth{6.0in}
\tabletypesize{\footnotesize}
\tablecaption{Newly Discovered Nuclear Masers.\label{Table}} \tablehead{
     \colhead{Galaxy}              &
     \colhead{Type\tablenotemark{(a)}}              &
     \colhead{$\alpha_{2000}$\tablenotemark{(b)}}      &
     \colhead{$\delta_{2000}$\tablenotemark{(b)}}   &
     \colhead{v$_{sys}$\tablenotemark{(c)}}    &
     \colhead{$\log_{10}\,L_{\mbox{\tiny H$_2$O}}$\tablenotemark{(d)}}    &
     \colhead{Date\tablenotemark{(e)}}           &
     \colhead{$T$\tablenotemark{(f)}}      &
     \colhead{$1\sigma$\tablenotemark{(g)}} &
     \colhead{Misc\tablenotemark{(h)}} \\
     \colhead{}             &
     \colhead{}             &
     \colhead{(hhmmss)}     &
     \colhead{(ddmmss)}     &
     \colhead{(km s$^{-1}$)} &
     \colhead{($\log_{10}\,L_{\odot}$)} &
     \colhead{(ddd-yy)}             &
     \colhead{(s)} &
     \colhead{(mJy)} &
     \colhead{} \\
}

\startdata

NGC\,0235A      & Sy2 & 00~42~52.81 & $-$23~32~27.7 & 6519  & 2.0    & 214-03 & 2280 & 15 & TV\\
                &     & 00~42~52.81 & $-$23~32~27.8 &       &        &        &      &    & BnA\\
NGC\,0613       & Sy  & 01~34~18.23 & $-$29~25~06.6 & 1468  & 1.2    & 174-03 & 1680 & 16 & TV\\
                &     & 01~34~18.18 & $-$29~25~06.5 &       &        &        &      &    & BnA\\
IC\,0184        & Sy2 & 01~59~51.23 & $-$06~50~25.4 & 5287  & 1.4    &183-05 & 8760 & 7  & GV\\
                &     & 01~59~51.23 & $-$06~50~25.4 &       &        &       &      &    & CnB\\
VII\,ZW\,073    & Sy2 & 06~30~25.57 & $+$63~40~41.2 & 11899 & 2.2    & 337-04 & 19560 & 6 & GRV\\
                &     & 06~30~25.54 & $+$63~40~41.3 &       &        & 353-04 &       &   & B\\
NGC\,3393       & Sy2 & 10~48~23.46 & $-$25~09~43.4 & 3704  & 2.4       & 167-03 & 6120 & 8 & GTV\\
                &     & 10~48~23.45 & $-$25~09~43.6 &       &           & 174-03 &      &   & BnA\\
NGC\,4293       & LINER & 12~21~12.89 &  $+$18~22~56.6 & 890  & 0.1-0.7 & 056-04 & 15420 & 5 & RV\\
                &       & 12~21~12.82 &  $+$18~22~57.4 &      &         & 063-04 &       &   & B\\
NGC\,5495       & Sy2 & 14~12~23.35 & $-$27~06~28.9 & 6589    & 2.3     & 215-03 & 2820 & 10 & GV\\
                &     & 14~12~23.35 & $-$27~06~29.2 &         &         &        &      &    & BnA\\
AM\,2158-380\,NED02  & Sy2 & 22~01~17.07 & $-$37~46~24.0 & 9661  & 2.7  & 175-05 & 11460 & 7 & GTV\\
                     &     & 22~01~17.10 & $-$37~46~23.0 &       &      & 178-05 &       &   & CnB\\
\enddata

\tablenotetext{(a)}{Activity type. NGC\,5495 is classified as Seyfert\,$2$ by
\cite{Kirhakos1990}, NGC\,3393, IC\,0184, VII\,ZW\,073, and AM\,2158-380\,NED02
are listed as Seyfert\,$2$ in the Veron-Cetty catalogue \citep{Veron2003},
NGC\,0613 is classified as a possible Seyfert by \cite{Veron1986}, NGC\,4293 is
listed as LINER by \cite{Ho1997b}, while NGC\,0235A is classified as Seyfert\,$1$
by \cite{Maia1987} but as Seyfert\,$2$ by \cite{Monk1986}. Refer to the text for
the discussion of the ambiguity in NGC\,0235A spectral classification.}

\tablenotetext{(b)}{{\it First line:} optical positions from the NED with
uncertainties of $\pm1\rlap{.}''3$. {\it Second line:} maser positions measured
with a VLA snapshot with typical uncertainties of $\pm0\rlap{.}''3$. In an effort
to test for systematic errors in the derived maser positions, we imaged disjoint
segments of the VLA data and confirmed that they yield consistent maser positions
(that is, within $0\rlap{.}''3$). }

\tablenotetext{(c)}{Heliocentric systemic velocity computed assuming the radio
definition of Doppler shift.}

\tablenotetext{(d)}{Total water maser luminosity assuming isotropic emission of
radiation and distances based on $H_o=75$\,km\,s$^{-1}$\,Mpc$^{-1}$, except for
for NGC\,0613 and NGC\,4293, whose distances were adopted to be $17.9$\,Mpc
\citep{Jungwiert1997} and $17$\,Mpc \citep{Tully1988}, respectively.}

\tablenotetext{(e)}{Data obtained at listed epochs were combined to form a single
spectrum.}

\tablenotetext{(f)}{Total integration time on+off\,source.}

\tablenotetext{(g)}{Rms noise in a $1.3$\,km\,s$^{-1}$ spectral channel,
corrected for atmospheric opacity (typically $\sim 0.07$) and for the dependence
of antenna gain on elevation.}

\tablenotetext{(h)}{{\it First line:} Instruments used to confirm the initial
detections: G$=$GBT, R$=$Robledo, T$=$Tidbinbilla, V$=$VLA. The velocities and
flux densities of the narrow spectral peaks in AM\,2158-380\,NED02, NGC\,0235A,
NGC\,3393, NGC\,5495, and NGC\,4293 detected with the VLA agree with those
measured using the Tidbinbilla antenna to within 1.3\,km\,s$^{-1}$ and $30\%$,
respectively. On the other hand, a $6$\,km\,s$^{-1}$ discrepancy in the velocity
centroid of the NGC\,0613 spectral-line likely reflects uncertainty due to noise
in the Tidbinbilla spectrum. Although the peak flux densities of the IC\,0184
maser agree to within $20\%$, the velocity centroids are in disagreement by
$5$\,km\,s$^{-1}$, which is again most likely due to low signal-to-noise ratio in
the Tidbinbilla spectrum. The strengths of the VII\,ZW\,073 maser measured with
the VLA and the Robledo antenna differ by a factor of $2.5$ (time baseline of
$5$\,months), which could be due to either maser variability or low
signal-to-noise ratio in both spectra. We note that such variability is not
unusual \citep[e.g.,][]{Baan1996}. The flux densities of AM\,2158-380\,NED02,
IC\,0184, NGC\,3393, NGC\,5495, and VII\,ZW\,073 measured with the DSN and the
GBT antennae agree to within 25\%. (The GBT spectra will be presented in a
follow-up article.) {\it Second line:} Configuration of the VLA. The VLA
bandwidth was $6.25$\,MHz (channel spacing of 1.3\,km\,s$^{-1}$) for all imaged
sources except NGC\,0613, where a wider bandwidth of $25$\,MHz (channel spacing
of 11\,km\,s$^{-1}$) was necessary in order to include all the emission.}

\end{deluxetable}

\begin{deluxetable}{llllrccrc}
%\tablewidth{6.0in}
\tablecaption{Sources Surveyed for Water Maser Emission with \hbox{$70$-m} Deep
Space Network Antennas at Tidbinbilla and Robledo. [Refer to the source for the
complete version of this table.]\label{Survey}} \tablehead{
     \colhead{Galaxy Name}              &
     \colhead{Type\tablenotemark{(a)}}              &
     \colhead{$\alpha_{2000}$\tablenotemark{(a)}}      &
     \colhead{$\delta_{2000}$\tablenotemark{(a)}}   &
     \colhead{v$_{sys}$\tablenotemark{(a)}}    &
     \colhead{Date}           &
     \colhead{$T$\tablenotemark{(b)}}      &
     \colhead{$1\sigma$\tablenotemark{(c)}} &
     \colhead{Site\tablenotemark{(d)}} \\
     \colhead{}             &
     \colhead{}             &
     \colhead{(hhmmss)}     &
     \colhead{(ddmmss)}     &
     \colhead{(km s$^{-1}$)}       &
     \colhead{}        &
     \colhead{(s)}          &
     \colhead{(mJy)}
}

\startdata

KUG2358+330 & SY2 & 00~00~58.14 &   $+$33~20~38.1 &         12387 & 2005-01-05 &         6240 &            9 & R \\
UGC12914 & LINER & 00~01~38.32 &   $+$23~29~01.5 &          4308 & 2003-02-03 &         2910 &           18 & R \\
UGC12915 & LINER & 00~01~41.94 &   $+$23~29~44.5 &          4274 & 2004-07-20 &         2460 &           12 & T \\
CGCG517-014 & SY2 & 00~01~58.46 &   $+$36~38~56.6 &          9311 & 2004-12-26 &         6060 &            9 & R \\
NGC7814 & LINER & 00~03~14.89 &   $+$16~08~43.5 &          1046 & 2002-08-15 &         1980 &           14 & T \\
UGC00013 & LINER & 00~03~29.23 &   $+$27~21~05.9 &          7498 & 2002-12-25 &        2100 &           23 & R \\

\enddata

\tablenotetext{(a)}{Type, position, and heliocentric systemic velocity obtained
from the NED at the outset of the survey in 2002. Velocities are computed
assuming the radio definition of Doppler shift.}

\tablenotetext{(b)}{Total integration time on+off\,source.}

\tablenotetext{(c)}{Rms noise in a $1.3$\,km\,s$^{-1}$ spectral channel,
corrected for atmospheric opacity (typically $\sim 0.07$) and for the dependence
of antenna gain on elevation.}

\tablenotetext{(d)}{R$=$Robledo, T$=$Tidbinbilla.}

\end{deluxetable}

\end{document}